# 400-Gbps/λ Ultrafast Silicon Microring Modulator for Scalable Optical Compute Interconnects

Fangchen Hu[1,3], Fengxin Yu[1,3], Xingyu Liu[2], Aoxue Wang[2], Xiao Hu[1,*], Haiwen Cai[1,*], and Wei Chu[1,*]

[1] *Zhangjiang Laboratory, Shanghai 201210, China*

[2] *Future Information Innovative College, Fudan University, Shanghai 200433, China*

[3] *These authors contribute equally.*

[*]*Corresponding author. E-mail: huxiao@zjlab.ac.cn (X.H.); caihw@zjlab.ac.cn (H.C.) and chuwei@zjlab.ac.cn (W.C.)*

**Abstract**

The exponential growth of artificial intelligence (AI) workloads is driving an urgent demand for optical interconnects with ultrahigh bandwidth, energy efficiency, and scalability. Silicon photonics, with its CMOS compatibility and wafer-scale manufacturability, has emerged as a promising platform for optical interconnect architectures. Silicon microring modulators (MRMs), with their compact footprint, low power consumption, and high modulation efficiency, have become ideal devices for modulation in interconnects. However, silicon MRMS have so far been constrained by the trade-off between modulation efficiency and bandwidth, hindering their potential for 400 Gbps-per-wavelength operation. To mitigate this trade-off, here we demonstrate a wafer-level fabricated and high-bandwidth silicon MRM with a novel heavily-doped trench-integrated structure on a 300-mm silicon photonic platform, achieving both outstanding device performance and remarkable wafer-scale uniformity. Exploiting dual operation modes: self-biasing for energy-efficient scale-up interconnects and depletion driving for ultrafast scale-out links, the device supports error-free 32-Gbps NRZ transmission over 2-km SSMF with only 0.43-Vpp drive and zero electrical bias, yielding energy efficiency of 0.97 fJ/bit without DSP. At higher swings, it further supports 280-Gbps PAM4 and error-free 80-Gbps NRZ optical modulation. For scale-out interconnects, open eye diagrams are achieved at 200 Gbps (NRZ), 360 Gbps (PAM4), and a record 400 Gbps (PAM6), establishing the first wafer-scale silicon MRM solution reaching 400 Gbps/λ. The sub-fJ/bit energy efficiency and high bandwidth demonstrated in this work establish the MRM as a scalable, high-performance solution for next-generation optical interconnect architectures in AI computing networks

**Keywords**: Silicon photonics, microring modulator, 400Gbps per wavelength, optical I/O, wafer-scale integration

# 1. Introduction

The rapid advancement of modern artificial intelligence (AI), exemplified by large language models (LLMs), multimodal systems, and generative architectures, has led to exponentially increasing computational demands for efficient model training and inference[1,2,3]. However, the slowdown of Moore's Law imposes fundamental constraints on the computational and memory capacities of individual processing units, such as central processing units (CPUs) and graphics processing units (GPUs)[4]. To overcome these limitations, data centers are increasingly adopting massively parallel computing architectures through both vertical (scale-up) and horizontal (scale-out) expansion of computing nodes (e.g. XPUs or XPU clusters)[5,6]. In parallel, resource pooling across compute, memory, and storage domains is being actively deployed to improve per-node memory availability and overall resources utilization[7]. As illustrated in Fig.1 a, both architectural scaling and resource disaggregation critically rely on a high-performance interconnect fabric, ideally one that is scalable in bandwidth, energy efficiency and transmission reach[8,9].

Traditional electrical interconnects, while widely deployed, increase input/output (I/O) bandwidth at the expense of link length, signal integrity, and power efficiency, thereby struggling to meet the demands of a scalable interconnect fabric. In contrast, silicon photonics (SiPh)-based optical interconnects, that is, a novel photonic fabric, have emerged as a highly competitive alternative, offering high bandwidth, long-reach fiber transmission, scalable multiplexing (e.g. wavelength division multiplexing, WDM), and complimentary-metal oxide-semiconductor (CMOS)-compatible mass manufacturing. These advantages enable photonic fabric to scale bandwidth cost-effectively with minimal penalties to transmission reach and signal fidelity[10]. Building on this foundation, the advent of co-packaged optics (CPO) further reduces SerDes-to-SiPh distances, substantially enhancing both energy efficiency and bandwidth density.[11] Among various implementations, microring-resonator (MRR)-based WDM architectures stand out, as MRRs integrate modulation and filtering in an ultracompact footprint (on an order of tens of micrometers) with low capacitance and minimal insertion loss, making it particularly suitable for constructing a scalable optical compute fabric[12,13,14].

As the key active components in MRR-based architectures, silicon microring modulators (MRMs) are being intensively investigated to meet the stringent requirements of both scale-up and scale-out. In scale-up scenarios, optical I/O (OIO) supports direct interconnection between XPUs or between XPUs and memory pools across racks in alignment with emerging standards such as Universal Chiplet Interconnect Express (UCIe) or Compute Express Link (CXL)[15]. These links demand not only high energy efficiency that matching or surpassing that of copper links, but also minimal reliance on digital signal processing (DSP) and forward error correction (FEC) overhead to enable reliable (error-free) and low-latency data transfer with bit error rate (BER) < 1e-12. To meet these requirements, recent efforts have focused on developing energy-efficient MRMs, as summarized in Fig.1e and Supplementary Table 1[16-23]. Most reported MRMs operate in depletion mode, leveraging the high carrier mobility and fast response of free-carrier plasma dispersion to maximize electro-optic

(EO) bandwidth. However, this approach inevitably sacrifices modulation efficiency due to the fundamental trade-off between EO bandwidth and modulation depth. To compensate for the reduced modulation efficiency, depletion-mode MRMs typically require large driving voltages (over 1 $V_{pp}$), which necessitates power-hungry driver circuits, increases circuit area, and raises overall system power dissipation[21]. Alternatively, depletion-mode athermal resonant silicon modulators have been proposed to enable low-$V_{pp}$ optical modulation (e.g. 0.5 $V_{pp}$)[17]. Yet, their electro-optic bandwidth remains limited to approximately 25 GHz, restricting the maximum data rate achievable per ring.

In scale-out scenarios, MRMs play a pivotal role in CPO transceivers for high-capacity ethernet-based switch interconnects between compute clusters, where higher per-wavelength line rates are essential for capacity scaling. The IEEE 802.3dj standard specifies 1.6 TbE based on 200 Gbps PAM4 intensity modulation/direct detection (IM/DD), while 400 Gbps-per-wavelength IM/DD solutions remain under active exploration (Fig.1f and Supplementary Table 2)[24-36]. While alternative photonic platforms, such as thin-film lithium-niobate Mach–Zehnder modulators (TFLN-MZM) and InP externally modulated lasers (InP-EML), have demonstrated 400 Gbps/λ capacities, their poor compatibility with CMOS fabrication and limited integration suitability for CPO present significant barriers to large-scale photonic–electronic co-integration. Similarly, resonance-enhanced SiPh modulators, including slow-light and plasmonic MRMs have achieved high speeds but suffer from extreme sensitivity to fabrication imperfections, which restricts their scalability for large-scale manufacturing.

Among SiPh solutions, silicon depletion-mode MRMs are particularly attractive due to their potential of high-speed characteristics approaching 110 GHz[32]. However, like their counterparts in scale-up scenarios, they remain fundamentally constrained by the bandwidth-efficiency trade-off, a common limitation across both scale-up and scale-out contexts.

To address these challenges, we present a novel Si MRM featuring a narrow trench structure with heavy doping, which synergistically alleviates the bandwidth-efficiency trade-off. The narrow trench broadens the optical bandwidth by introducing propagation loss, while reduced series resistance from this design extends electrical bandwidth. Heavy doping further enhances modulation efficiency by increasing carrier density. Fabricated on a 300 mm silicon photonic platform, devices within a wafer achieve an electro-optic bandwidth exceeding 110 GHz at −3 V bias and 80 GHz at 0 V bias (without optical peaking), along with a modulation efficiency V$\pi$·L of ~0.57 V·cm. Benefitting its outstanding performance in bandwidth and efficiency, we propose a dual operation modes of the device, depletion and self-biasing mode, to address scale-out demands at 400 GbE and scale-up requirements on modulators with sub-fJ/bit energy efficiency, respectively. In UCIe-based XPU scale-up scenarios, the self-biasing MRM achieves error-free 32 Gbps NRZ transmission over 2 km standard single-mode fiber (SSMF) with only 0.43 Vpp drive and zero electrical bias, without requiring any DSP, resulting in an energy consumption of merely 0.97 fJ/bit. Higher data rates up to 280 Gbps PAM4 and error-free 80 Gbps NRZ are further supported with modest penalties in drive swing, DSP, and optical power. The elimination of bias-tee and driver circuits substantially reduces the energy and footprint of massively parallel photonic engines[37], yielding a sub-fJ/bit MRM fully compliant with the voltage swing requirement of UCIe 2.0 (~0.4 Vpp). For scale-out interconnects, we demonstrate open eye diagrams at several record-high data rates, including 200 Gbps (NRZ), 360 Gbps (PAM4), and 400 Gbps (PAM6). Our results presented in Fig. 1e-f and Extended Data Table 1 demonstrate state-of-the-art performance, surpassing prior benchmarks in key metrics. This work thus presents a novel, manufacturable 400 Gbps microring modulator solution that supports the sustained expansion of next-generation AI computing networks, while underscoring the enduring potential of silicon photonics to deliver scalable bandwidth beyond 100 GHz.

## 2. Results

**Device design**

The EO bandwidth ($f_{EO}$) of silicon MRMs is fundamentally governed by three factors: the optical bandwidth ($f_{ph}$), determined by photon lifetimes (or equivalently, the quality factor Q) of optical mode, the electrical bandwidth ($f_e$), determined by RC time constants of the PN junction and metal contacts, and the wavelength detuning from resonance[38,39]. Although detuning the resonance can exploit optical peaking to extend $f_{EO}$, it will significantly degrade modulation efficiency by shifting the operating point away from optimal extinction. We therefore focus on maximizing $f_{ph}$ and $f_e$ while maintaining modulation efficiency at a practical level. Their relationship is formally expressed as[40]:

$$\frac{1}{f_{EO}^2} = \frac{1}{f_{ph}^2} + \frac{1}{f_e^2} \quad (1)$$

Here, the photon lifetime $\tau_{ph}$ is directly related to the cavity quality factor $Q$ through $Q = \omega\tau_{ph}$, where $\omega$ is the angular frequency. Under critical coupling (coupling loss equals propagation loss), the photon decay length is $(2\alpha)^{-1}$, and with $v_g = c/n_g$ ($n_g$: group index), the photon-lifetime-limited bandwidth[41] can be expressed as:

$$f_{ph} = \frac{1}{2\pi\tau_{ph}} = \frac{c\alpha}{n_g\pi} \approx \frac{c}{\lambda Q} \quad (2)$$

where $c$ is the light speed, $\alpha$ is the optical power loss coefficient, and $\lambda$ is the resonance wavelength. The RC-limited electrical bandwidth is generally expressed as:

$$f_e = \frac{1}{2\pi(R_s + R_{dr}) \times C_j} \quad (3)$$

where $C_j$ is the PN junction capacitance, $R_s$ is the modulator series resistance and driver impedance is $R_{dr}$. The mutual constraints between optical and electrical bandwidths in conventional MRMs hinder simultaneous optimization. By employing a narrow-trench geometry, we introduce additional design freedom that decouples the two bandwidths and enables a substantial increase in EO bandwidth, while heavy doping ensures that modulation efficiency remains at a practical level.

Fig. 1a depicts the cross-section of our novel heavily-doped low-Q silicon MRM with narrow trench structure. Unlike traditional designs (shown in Supplementary Note 1.A), our proposed MRM features a narrow trench and high doping concentration in rib waveguide. This configuration induces additional optical loss in the ring, resulting in reduced Q-factor and significantly increased optical bandwidth. Narrow trench geometries coupled with heavy doping can also reduce series resistance, boosting RC-governed electrical bandwidth. Furthermore, high doping concentration at the P-N junction interface enhances modulation efficiency and spectral symmetry at 0 V bias[42], enabling distortion-free operation under zero-bias conditions.

Fig. 1b is a schematic of our proposed silicon MRM. The present device is designed for 220-nm-high silicon-on-insulator (SOI) platform. It consists of a 420 nm-wide, 220nm-thick Si rib waveguide with a 70-nm-thick slab, connected to another 220-nm-thick slab. The 220-nm-thick slab inside and outside the ring are designed to form two trench structure for extra optical loss, which can significantly reduce photon lifetime and increase photon lifetime-limited bandwidth. To achieve critical coupling without exciting other higher-order modes, the coupling region length and gap width are carefully designed. A bus waveguides were used in the coupling region to enhance coupling with the resonator. Further details of coupling region design in MRM are provided in Supplementary Note 1.B. The cross-section of the MRM was characterized using a focused ion beam-scanning electron microscope (FIB-SEM) (Fig. 1c), which showed a ring rib waveguide with a radius of 8 μm, tungsten heater, PN junction, vias and metals. The rib waveguide enables a simple lateral P-N junction cross section design, which is depicted by the inset of Fig. 1c. The P-N junction is accessed through highly-doped 70-nm-thick silicon slab and ohmic contacts with vias and metals on 220-nm-thick slab. The inner ring rail is fabricated from phosphorus-doped silicon, while the outer rail consists of boron-doped silicon. The proposed MRM is fabricated in a 12-inch CMOS foundry. The photograph of the fabricated 300-mm wafer and the microscope picture of silicon MRM are shown in Fig. 1d.

The structure of P-N junction cross section in MRM can be optimized to have an extremely high bandwidth by adjusting both the trench width and doping concentration in rib waveguide. In our design, Q factors and optical bandwidth of MRM variation against the trench width and doping concentration are simulated (based on equation 2 and Supplementary Note 1.C. for details), as shown in Fig. 2a and Fig. 2e. The corresponding Q values are calculated under the situation that all resonators are critically coupled to the bus waveguides. As the trench narrows and doping concentration increases, waveguide loss rises and the Q-factor degrades, significantly broadening the optical bandwidth. Compared to the low-doping traditional structure without trench structure, the narrow-trench/high-doping design enhances optical bandwidth from 50 GHz to over 200 GHz. To further investigate trench width and doping concentration effects on electrical bandwidth ($f_e$), we simulate maps of junction capacitance ($C$) and series resistance ($R$) on varying trench width and doping-level at -3 V (illustrated in Fig. 2b-c) and 0 V (Fig. S3c-d in Supplementary Note 1.C). Further details and results of capacitance and resistance simulation are provided in Supplementary Note 1.B. We can observe that the doping level and trench width of the PN junction would influence such two attributes in a contrary way. As presented in Fig. 2b-c, a higher doping concentration and smaller trench width demote the series resistance, but the junction capacitance also increases with heavier doping and narrower trench. Based on these results, when the modulation length is set 40μm to achieve sufficient modulation efficiency, the electrical bandwidth at -3 V (Fig. 2 d) and 0 V (Fig. S2e) is calculated by equation 3. At -3 V, compared to the traditional structure, the narrow-trench/high-doping design can enhance electrical bandwidth from 80 GHz to over 110 GHz. Finally, by integrating optical and electrical bandwidth via Equation. 1, we plot the total EO bandwidth of the MRM versus trench width and doping concentration in Fig. 2f. It is obvious that narrow trench can significantly boost the EO bandwidth to over 100GHz. Higher doping level can further enhance the EO bandwidth to over 80 GHz under zero bias operation (Fig. S2f in Supplementary Note 1B). Therefore, to support the 400 Gbps/λ data transmission, we choose a relatively narrow trench width and high doping level for our design, which gives $f_{EO}$ ~ 100 GHz at -3 V and 85 GHz at 0 V.

**D.C. and EO characteristics**

To evaluate and assess the performance of the proposed design, measurements were conducted on a fabricated MRM. The measured transmission spectrums of a single MRM (represented by the blue curve) and a single pair of grating couplers (represented by the orange curve) under low power condition are presented in Fig. 3a. Comparison of the two curves yields an average off-resonance insertion loss of 1 dB for the MRM. The MRM exhibits a DC ER of ~49 dB, a full-width at half-maximum (FWHM) of ~0.88 nm and a free spectral range (FSR) of 7 nm. A Q-factor of 1500 is extracted from the transmission spectrum of the MRM. Normalized optical spectra with different bias voltages applied, from −4 V to 1 V, are shown in Fig. 3b. Regardless of bias direction (forward/reverse), ER diminishes, confirming critical coupling operation at 0 V for this MRM. From +1V to -4V, the resonance wavelength shifts to the red side and the coupling condition changed among over-, critical- and under-coupling. The average wavelength shift (Δλ/ΔV) reaches 24.5 pm/V as the bias voltage varies from 0 to -4 V.

The figure of merit for modulation efficiency, Vπ·L, then can be calculated based on $V_\pi L = \frac{FSR \times L}{2(\Delta\lambda/\Delta V)}$ , where L is the length of doping region. Therefore, the Vπ·L of the proposed MRM is ~0.57 V·cm. Owing to high doping concentration in the MRM, ER and Δλ/ΔV variation is small when forward bias below 0.8 V, contributing to very small signal modulation distortion at zero bias operation. Each MRM uses an on-chip tungsten heater to control its resonance wavelength. The MRM transmission spectrum versus the tungsten heater bias voltage, spanning from 0 V to 2 V, is plotted in Fig. 3c. The integrated tungsten heater has a resistance of ~54 Ω. The resonance red-shifts as voltage raises, and the corresponding resonance wavelength against heater power is shown in Fig. 3c, where the red-shift is ~81.2 pm/mW. The DC consistent performance exhibited by the MRM was assessed through spectrum measurements on nine distinct dies from the same 12-inch wafer and the results are provided in Fig. S4 in Supplementary Note 2. A. Consistent ER, Q-factor, $V_\pi L$, and heater efficiency across nine dies demonstrate the MRM's suitability for mass production.

The device was characterized at high speed by measuring the EO S21 using a lightwave component analyser (LCA) through a GS probe. Beside photon lifetime bandwidth and RC bandwidth mentioned before, wavelength detuning also influence the EO bandwidth. The MRM achieves enhanced 3-dB bandwidth at wavelengths detuned from resonance through optical peaking. However, the Lorentzian line-shape of its transmission spectrum exhibits a nonlinear wavelength-dependent slope. Such bandwidth enhancement inevitably attenuates low-frequency response amplitudes and degrades modulation efficiency. This reduction in modulation slope fundamentally limits any potential improvement to OMA and ER in eye diagrams via optical peaking. Fig. 3d shows the EO S21 at 0 V and -3 V under different detuning wavelengths (located at the 4 dB, 5 dB and 6 dB insertion loss (IL) point on the transmission spectrum). At 0 V, the measured EO bandwidths are 83 GHz, 97 GHz and >110 GHz at 6 dB, 5 dB and 4 dB operating point. At -3 V, the all measured EO bandwidths are >110 GHz at 6 dB, 5 dB and 4 dB operating point. With the increase of the reversed-bias voltage, the bandwidth also increases due to the reduction of the P-N junction capacitance. The bandwidth values align with the simulated EO bandwidth obtained from the RC-limited bandwidth and photon lifetime-limited bandwidth. To demonstrate the consistent EO performance of the MRM, bandwidths were measured across nine different dies on a single 12-inch wafer at different operating points. At 6 dB operating point, as depicted in Fig. 3e-f, all of these MRMs exhibit > 82 GHz bandwidth at 0V and > 108 GHz bandwidth at -3V. Exceptional bandwidth uniformity enables commercial-scale production and deployment of high-speed MRMs. More S21 curves at different operating points from nine dies are shown in Fig. S5 in Supplementary Note 2. B.

**High-speed transmission performance**

OIO is optimized for AI infrastructure (e.g. XPUs) to maximize compute efficiency while reducing costs, latency, and power consumption. Unlike traditional bulky and power-hungry pluggable optics, OIO is integrated into the interposer of XPU package, aiming to eliminate communication bottlenecks of copper interconnection and enable large-scale compute node interconnection. Figure 4(a) illustrates the architecture of an OIO module for interconnecting two XPUs. The electrical interface between the XPU and the OIO employs die-to-die (D2D) links compliant with UCIe specification to achieve high electrical edge interconnection density. A single module configuration of UCIe 2.0 supports x8 or x16 data interfaces with edge bandwidth density up to 1.79 Tb/s/mm, energy efficiency with 1.25 pJ/bit and link rates up to 32 GT/s that aligned with the PCI Express (PCIe) specifications.

In the OIO transmitter (OIO Tx), multi-lane UCIe signals from XPUs are directly or retimed before being amplified and biased by a packaged electrical integrated chip (EIC), and then fed into the microring modulator (MRM) array. On the receiver side, 8 or 16 mirroring resonators are used to demultiplex the optical signals, which are subsequently amplified using transimpedance amplifiers (TIAs) and variable-gain amplifiers (VGAs), and delivered to the target XPU via UCIe-compliant D2D interfaces. Within the transmitter, the EIC dominates area density and power consumption compared to the photonic integrated circuit (PIC)[19,21,43]. For example, the combined power of the DC bias, driver, and VGA for a 64 Gbaud NRZ MRM can reach ~81 mW, whereas the laser power is only ~5 mW[21]. Moreover, the spacing between adjacent MRMs is primarily constrained by bump-induced electrical crosstalk rather than the physical size of the rings[19]. To address these limitations, we propose a bias-free and driverless transmitter based on our MRM operated at self-biasing mode, aiming to maximize both energy efficiency and area bandwidth density in the OIO transmitter.

To demonstrate the compatibility of this architecture with the lane speed of PCIe standards, we first experimentally measured eye diagrams at rates of PCIe 5.0 to PCIe 7.0 using the setup shown in Fig. 4b. The peak-to-peak voltage ($V_{pp}$) of the RF signal driving the MRM was swept from 0.28 V to 0.72 V, which is below both UCIe 2.0 voltage swing specification (0.4 V) and the core voltage (0.87 V) of a 12 nm FinFET CMOS process, respectively [44,45]. The low voltage swing operation ensures tight voltage compatibility with modern CMOS systems and eliminates the need for additional bias or drive circuitry. Fig.4 (c-d) show the optical eye heights and eye width for PCIe 5.0 signals as a function of $V_{pp}$ and optical received power (Rop). The measured eye width meets the requirement of PCIe 5.0 (9.375 ps). The measured optical eye heights provide guidance for selecting TIAs and VGAs to further meet the requirement of electrical eye height. For example, we can select a 79 Ω TIA operating at the Nyquist point of a 32G NRZ signal[46] to satisfy the 15 mV eye height requirement of PCIe 5.0 when $V_{pp}$ = 0.43 V and Rop = 5 dBm. if using TIA and VGA, the requirement on Rop and $V_{pp}$ can be further relaxed. Eye analysis on PCIe 6.0 is provided in the Fig. S7 in Supplementary Note 3. A. Representative eye diagrams for PCIe 5.0, 6.0 and 7.0 are given in Fig. 4e-g. NRZ and PAM4 signals are generated using PRBS11-1 (8192 symbols) and SSPRQ (66535 symbols) patterns, respectively. The BER of PCIe 5.0 optical eye diagram at Vpp = 0.43 V and Rop = 3 dBm is lower than 1e-12 without using any digital signal processing (DSP). When applying 3-taps FFE, up to 80 Gbps NRZ signal can be error-free transmitted over 2km SSMF (Fig. S8 in Supplementary Note 3. B.). The transmitter dispersion eye closure quaternary (TDECQ) of the PCIe 6.0 and 7.0 signals are 1.87 and 2.66 dB, respectively, measured at the Rop = 5 dBm. TDECQ was evaluated under a soft-decision forward error correction (SD-FEC) threshold of symbol error rate (SER) at 1E-2.

To assess the scalability of our device beyond PCIe 7.0[48], higher speed signals were tested using a 16-tap feed-forward equalizer (FFE) and increasing $V_{pp}$ to 0.73 V. The number of taps is modified for best eye diagram quality analyzed in Fig. S9 in Supplementary Note 3. C. Eye diagrams for 180 Gbaud NRZ and 140 Gbaud PAM4 signals are shown in Figs. 4h and 4j, respectively. Although the PCIe 8.0 standard is not yet finalized, Fig. 4 (i) demonstrates the potential for future support using the proposed MRM-based optical link at the penalty of more complex DSP. The TDECQ of a 256 GT/s PAM4 signal is 3.62 dB after equalization. According to the power analysis in the Methods section, the $E_{bit}$ values for the PCIe 5.0–7.0 are 0.97, 0.49 and 0.78 pJ/bit, respectively.

With the evolution of co-packaged optics (CPO) architectures (Fig. 5a), the electrical interconnect length between switch ASICs and optical engines continues to shrink, driven by the push for higher single-wavelength data rates beyond 200 Gbps. This reduction in interconnect distance—enabled by die-to-die integration and flip-chip bonding—has made compact, high-bandwidth-density transceivers increasingly critical. In this context, our depletion-mode MRM offers a highly compact and energy-efficient solution capable of supporting 400 Gbps single-wavelength transmission, addressing the growing demand for high-throughput switch-to-switch interconnects. In the illustrated architecture, the electrical integrated circuit (EIC) is flip-chip bonded directly atop the photonic integrated circuit (PIC) to minimize interconnect parasitic. High-speed signals generated by the ASIC's SerDes are routed through high-speed substrates to the EIC, which re-drives them to the MRMs via micro-bumps. The MRMs are operated in reverse-biased (depletion) mode and driven with large swing voltages to ensure high extinction ratios required for ultrafast optical modulation.

Figure 5b presents the experimental setup used to evaluate the MRM's high-speed optical modulation capabilities. The device was biased at –3 V and driven by a 2 $V_{pp}$ RF signal. While the MRM itself exhibits an electro-optic bandwidth exceeding 100 GHz, the bandwidth of the arbitrary waveform generator (AWG) used here was limited to ~55 GHz, resulting in high-frequency signal degradation. Accordingly, digital signal processing (DSP) was employed at both transmitter and receiver to compensate for the instrumentation bottleneck and to demonstrate the modulator's true potential. We conducted comprehensive eye diagram measurements using various modulation formats—including NRZ, PAM4, and PAM6—under both back-to-back (BtB) and 2 km standard single-mode fiber (SSMF) transmission conditions, as shown in Fig. 5(c). Open optical eyes were achieved at 200 Gbaud NRZ (SNR ~3.2 dB) and 180 Gbaud PAM4 (TDECQ ~3.4 dB). Even at 400 Gbps PAM6, open eye diagrams were obtained at received optical powers around 5.75 dBm. These results highlight the MRM's suitability for ultrahigh-speed interconnects, with its performance ultimately limited by test equipment bandwidth rather than intrinsic device speed. To further isolate the impact of the AWG, we also captured electrical eye diagrams by directly connecting the AWG to a digital communication analyzer (DCA), as detailed in Supplementary Note 3. D. The similarity between electrical and optical eyes confirms the MRM's capability to support next-generation ultrahigh-capacity optical interconnects for large-scale AI network scale-out.

## 3. Discussion

By co-optimizing modulation efficiency, optical bandwidth, and electrical bandwidth via a novel heavily-doped narrow trench structure, we realize a high-performance silicon MRM that breaks the bandwidth–efficiency trade-off, marking a significant advancement over previously reported silicon depletion mode MRMs. This represents the first realization of a SiPh modulator with an EO bandwidth exceeding 110 GHz on a 300 mm SiPh platform without pronounced peaking effects, while simultaneously maintaining a practical modulation efficiency of $V_\pi \cdot L \approx 0.57$ V·cm. In addition, the device exhibits excellent wafer-scale uniformity, with a bandwidth variation within ~10 GHz, and maintains a high modulation efficiency with a $V_\pi \cdot L$ less than 0.6 V·cm. Based on the outstanding performance of our device, this work dispels the prevailing concern that silicon modulators cannot support 400 Gbps-per-wavelength operation, demonstrating that CMOS-compatible MRMs can rival emerging platforms such as TFLN and InP. Beyond scale-out interconnects, the same device also enables energy-efficient optical-I/O transmitters in self-biasing mode, underscoring its versatility for both high-speed and low-power applications.

Owing to its inherent WDM compatibility, the ultrafast MRM provides a highly scalable foundation for both compute node scale-up and AI network scale-out, supporting up to 2 Tbps and 2.4 Tbps interconnect capacities, respectively, as illustrated in Fig. S12. For a WDM OIO transmitter operating at 64 Gbaud PAM4 per wavelength, the required channel spacing must exceed 64 GHz to suppress inter-channel crosstalk. By allocating a 76 GHz spacing (including a 13 GHz guard band), the full free spectral range (FSR) of the MRM can accommodate 16 wavelength channels, enabling an edge bandwidth density of 8 Tbps/mm with a 0.25 μm fiber pitch. For a WDM CPO transceiver targeting 400 Gbps per wavelength using 155 Gbaud PAM6 signaling, six channels with 200 GHz spacing (including 45 GHz guard bands) can be supported. By reducing the microring radius to 4.5 μm, the FSR can accommodate eight 400 Gbps channels, laying the groundwork for next-generation 3.2 Tbps CPO transceivers. Altogether, this work establishes a silicon microring modulator platform that not only delivers record-breaking per-channel performance, but also offers a practical and scalable roadmap for terabit-class AI computing and datacenter interconnects.

**Materials and Methods**

**Fabrication**

The designed silicon MRM chips were fabricated using a standard foundry process on a 12-inch 220-nm-thick silicon-on-insulator wafer. The PN junction was formed from Phosphorous (for n-type) and Boron (for p-type) implants with 2 doping layers for the waveguide core to boost the optical bandwidth. The uniformity of the silicon MRM chips is shown in *Supplementary Note2* Fig. S4 and S5.

**Simulation**

The doping profile of the PN junction in MRM was simulated by Synopsys Sentaurus Process. The series resistance and junction capacitance were obtained with Synopsys Sentaurus Device. The optical field, electrical field and transmission spectrum of the MRM were obtained with the Lumerical FDTD, CHARGE and INTERCONNECT solver.

**Bandwidth characterization**

Bandwidth characterization employed a Keysight N5290A PNA MM-Wave System with N4372E 110-GHz Lightwave Component Analyzer to measure linear EO transmission. RF interconnects cables and bias-tee were precalibrated to 110 GHz prior to testing. During bandwidth characterization, coupled optical power into the bus waveguide was limited to 500 µW. This suppresses resonator nonlinearities — specifically two-photon absorption, free-carrier absorption, and self-heating — that otherwise induce resonance drift and would require precise thermal control for operating-point stability. The light after the output grating coupler was amplifier by a Praseodymium-doped fiber amplifier (PDFA) to keep the optical power entering the PD at 1 mW.

**Eye diagrams measurements**

Eye diagram measurements of the MRMs were performed using a 224 GSa/s arbitrary waveform generator (AWG, Keysight M8199B) with a nominal analog bandwidth exceeding 80 GHz. The AWG provided PRBS11-1 NRZ and SSPRQ PAM4 signals to drive the MRM. A 16 GHz reference clock, generated by a clock source (Keysight M8008A), was used to trigger both the AWG and a digital communication analyzer (DCA). Due to high-frequency degradation in RF cables and connectors, the actual electrical bandwidth of the direct link between the AWG and the DCA was limited to ~55 GHz. Similarly, although the AWG supports a maximum voltage swing of 2.7 V, the actual swing applied to the device under test (DUT) was attenuated by cable losses and estimated using a simple attenuation factor, detailed in the Supplementary Note 3.

D. To apply both RF and DC bias signals, a 110 GHz bias-tee was employed. A commercial tunable laser (EXFO, T200S-O) with a nominal output power of 10 dBm across the O-band served as the optical carrier. The modulated optical signal from the MRM was amplified by a praseodymium-doped fiber amplifier (PDFA, Thorlabs PDFA100) to compensate for link losses, primarily originating from ~9 dB insertion loss of two grating couplers. The MRM worked at 6 dB operating point. The optical output was received by a 90 GHz optical module (Keysight N1032A) on the DCA. Signal distortion introduced by electrical components was compensated through internal calibration tools.

To suppress amplified spontaneous emission (ASE) noise from the PDFA, signal waveforms were averaged over 16 acquisitions. A half-baud rate fourth-order Bessel low-pass filter and appropriate feedforward equalization (FFE) were applied at the receiver side. The transmitter dispersion eye closure quaternary (TDECQ) of PAM4 signals was evaluated at a symbol error rate (SER) threshold of 1e–2, consistent with SD-FEC requirements. Due to the limited bandwidth of the measurement setup, eye openings at 155 Gbaud PAM4 were barely discernible, as shown in the Fig.S9 of Supplementary Note 3. E. It's expected to get a clearer eye diagram for 400 Gbps PAM6 signal with higher-bandwidth AWGs and improved RF components.

**Modulator energy efficiency analysis**

Electrical power in the proposed device is primarily dissipated during rising transitions by charging the junction capacitance $C_j$ of the MRM. Assuming equidistant voltage levels in a given PAM-$N$ modulation scheme, the total consumed energy is expressed

as

$$E_p = C_j V_{pp}^2 \sum_{i=1}^{N-1}(N-i)(\frac{i}{N-1})^2,$$

where $V_{pp}$ is the peak-to-peak voltage swing and $N$ is the PAM order. Having $N^2$ total possible transitions and $\log_2(N)$ bits per symbol, the averaged energy consumed per bit is then given by

$$E_b = \frac{E_p}{N^2 \log_2^N}$$

The junction capacitance $C_j$ of out MRM is estimated to be 21 fF under zero electrical-bias and decreased to 16 fF under a -3 V electrical-bias. Under zero electrical-bias, the effective energy consumed per bit for modulation is calculated to be 0.97 fJ/bit (at 0.43 $V_{pp}$) for NRZ and 0.78 fJ/bit (at 0.73 $V_{pp}$) for PAM4. Under -3 V electrical-bias, the respective values rise to 16 fJ/bit (2 $V_{pp}$) for NRZ, 4.44 fJ/bit for PAM4, and 2.89 fJ/bit for PAM6. Note that only the energy efficiency of modulators is considered in these calculations.

At the transmitter level, the proposed OIO architecture eliminates the additional DC bias and electrical drivers, so the total transmitter power solely consists of contributions from the laser, PDFA and MRM. The optical power requirement depends on coupling loss $\alpha_i$, optical modulation bias of MRM $\alpha_{bias}$ and the required received optical power $P_{Rx}$. Assuming a wall-plug efficiency $\beta = 18\%$ for both laser and PDFA18 and a data rate $DR$, the overall energy efficiency of the OIO transmitters $E_{tx}$ can be estimated as:

$$E_{tx} = \frac{DR}{(P_{Rx} + 2 \times \alpha_i + \alpha_{bias})/\beta} + E_b$$

Based on this model, the $E_{tx}$ corresponding to the data rates of PCIe 5.0, 6.0, and

7.0 shown in Fig. 4(e–g) are estimated to be 1.96 pJ/bit, 1.56 pJ/bit, and 0.78 pJ/bit, respectively. These values exclude the coupling loss from grating couplers, which can be significantly reduced to $\alpha_i = 1$ dB/facet using modified edge couplers[19].

**References:**

1. S. Naffziger, "Innovations For Energy Efficient Generative AI," in *2023 International Electron Devices Meeting (IEDM)*, 2023, pp. 1-4.

2. P. Villalobos, J. Sevilla, T. Besiroglu, L. Heim, A. Ho, M. Hobbhahn, "Machine Learning Model Sizes and the Parameter Gap," arXiv: 2207.02852.

3. G. Wang *et al.*, "Zero++: Extremely efficient collective communication for giant model training," *arXiv preprint arXiv:2306.10209,* 2023.

4. Sevilla, J. et al. Compute trends across three eras of machine learning. In *2022 International Joint Conference on Neural Networks (IJCNN)* 1–8 (IEEE, 2022).

5. Y. Kundu *et al.*, "A Comparison of the Cerebras Wafer-Scale Integration Technology with Nvidia GPU-based Systems for Artificial Intelligence," *arXiv preprint arXiv:2503.11698,* 2025.

6. S. Amiralizadeh and J. K. Doylend, "AI Networking Challenges – A System Perspective," *IEEE Journal of Selected Topics in Quantum Electronics,* vol. 31, no. 3: AI/ML Integrated Opto-electronics, pp. 1-7, 2025.

7. K. Lim, J. Chang, T. Mudge, P. Ranganathan, S. K. Reinhardt, and T. F. Wenisch, "Disaggregated memory for expansion and sharing in blade servers," in Proc. 36th Annu. Int. Symp. Comput. Archit. New York, NY, USA: Association for Computing Machinery, Jun. 2009, pp. 267–278.

8. R. Lin, Y. Cheng, M. D. Andrade, L. Wosinska, and J. Chen, "Disaggregated Data Centers: Challenges and Trade-offs," IEEE Communications Magazine, vol. 58, no. 2, pp. 20-26, 2020.

9. Q. Cheng, M. Bahadori, M. Glick, S. Rumley, and K. Bergman, "Recent advances in optical technologies for data centers: a review," *Optica,* vol. 5, no. 11, pp. 1354-1370, 2018/11/20 2018.

10. S. Shekhar *et al.*, "Roadmapping the next generation of silicon photonics," *Nature Communications,* vol. 15, no. 1, p. 751, 2024/01/25 2024.

11. Mahajan, R. et al. Co-packaged photonics for high performance computing: status, challenges and opportunities. J. Lightwave Technol. 40, 379–392 (2021).

12. S. Daudlin *et al.*, "Three-dimensional photonic integration for ultra-low-energy, high-bandwidth interchip data links," *Nature Photonics,* 2025/03/21 2025.

13. M. Eppenberger *et al.*, "Resonant plasmonic micro-racetrack modulators with high bandwidth and high temperature tolerance," *Nature Photonics,* vol. 17, no. 4, pp. 360-367, 2023/04/01 2023.

14. Y. Peng et al., "An $8 \times 160$ Gb s$-1$ all-silicon avalanche photodiode chip," Nature Photonics, vol. 18, no. 9, pp. 928-934, 2024/09/01 2024.

15. V. Jain *et al.*, "Design Approach for Die-to-Die Interfaces to Enable Energy-Efficient Chiplet Systems," presented at the Proceedings of the 29th ACM/IEEE International Symposium on Low Power Electronics and Design, Newport Beach, CA, USA, 2024. [Online].

16. A. Rizzo *et al.*, "Massively scalable Kerr comb-driven silicon photonic link," *Nature Photonics,* 2023/06/29 2023.

17. E. Timurdogan, C. M. Sorace-Agaskar, J. Sun, E. Shah Hosseini, A. Biberman, and M. R. Watts, "An ultralow power athermal silicon modulator," *Nature Communications,* vol. 5, no. 1, p. 4008, 2014/06/11 2014.

18. Koch, B. R. et al. Integrated silicon photonic laser sources for telecom and datacom. In Optical Fiber Communication Conference, PDP5C–8 (Optica Publishing Group, 2013).

19. S. Daudlin *et al.*, "Three-dimensional photonic integration for ultra-low-energy, high-bandwidth interchip data links," *Nature Photonics,* 2025/03/21 2025.

20. C. Xie *et al.*, "A 64 Gb/s NRZ O-Band Ring Modulator with 3.2 THz FSR for DWDM Applications," in *2024 Optical Fiber Communications Conference and Exhibition (OFC)*, 2024, pp. 1-3.

21. N. Qi *et al.*, "A Monolithically Integrated DWDM Si-Photonics Transceiver for Chiplet Optical I/O," *IEEE Journal of Solid-State Circuits,* pp. 1-13, 2025.

22. C. Sun, "Photonics for Die-to-Die Interconnects: Links and Optical I/O Chiplets", in ISSCC 2024, Forum 1.7.

23. M. Wade *et al.*, "An Error-free 1 Tbps WDM Optical I/O Chiplet and Multi-wavelength Multi-port Laser," in *2021 Optical Fiber Communications Conference and Exhibition (OFC)*, 2021, pp. 1-3.

24. Y. Zhang et al., "240 Gb/s optical transmission based on an ultrafast silicon microring modulator," Photon. Res., vol. 10, no. 4, pp. 1127-1133, 2022/04/01 2022.

25. Hu, F. et al. 300-Gbps optical interconnection using neural-network based silicon microring modulator. Communications Engineering 2, 67, doi:10.1038/s44172-023-00115-x (2023).

26. S. Zhao et al., "High Bandwidth and Low Driving Voltage Add-Drop Micro-Ring Modulator for Optical Interconnection I/O Chips," Journal of Lightwave Technology, pp. 1-9, 2025.

27. X. Wang et al., "A 290 Gbps Silicon Photonic Microring Modulator with 83 -aJ/bit Power Consumption," in Optical Fiber Communication Conference (OFC) 2025, San Francisco,

California, 2025: Optica Publishing Group, p. M3K.5.

28. Y. Yuan et al., "A 5 × 200 Gbps microring modulator silicon chip empowered by two-segment Z-shape junctions," Nature Communications, vol. 15, no. 1, p. 918, 2024/01/31 2024.

29. D. W. U. Chan, X. Wu, C. Lu, A. P. T. Lau, and H. K. Tsang, "Efficient 330-Gb/s PAM-8 modulation using silicon microring modulators," Opt. Lett., vol. 48, no. 4, pp. 1036-1039, 2023/02/15 2023.

30. M. Sakib et al., "A 240 Gb/s PAM4 Silicon Micro-Ring Optical Modulator," in 2022 Optical Fiber Communications Conference and Exhibition (OFC), 2022, pp. 01-03.

31. D. W. U. Chan, X. Wu, Z. Zhang, C. Lu, A. P. T. Lau, and H. K. Tsang, "Ultra-Wide Free-Spectral-Range Silicon Microring Modulator for High Capacity WDM," Journal of Lightwave Technology, vol. 40, no. 24, pp. 7848-7855, 2022.

32. K. Lu, H. Chen, W. Zhou, H. K. Tsang, and Y. Tong, "Whispering Gallery Mode Enhanced Broadband and High-Speed Silicon Microring Modulator," Journal of Lightwave Technology, pp. 1-7, 2025.

33. Xin Chen, et al., "540Gbps IMDD Transmission over 30km SMF using 110GHz Bandwidth InP EML" Optical Fiber Communication Conference (OFC), Th4B.2, 2025.

34. Berikaa, E. *et al.* TFLN MZMs and Next-Gen DACs: Enabling Beyond 400 Gbps IMDD O-Band and C-Band Transmission. *IEEE Photonics Technology Letters* **35**, 850-853, doi:10.1109/LPT.2023.3285881 (2023).

35. Han, C. *et al.* Exploring 400 Gbps/λ and beyond with AI-accelerated silicon photonic slow-light technology. *Nature Communications* **16**, 6547, doi:10.1038/s41467-025-61933-5 (2025).

36. Shen, J. *et al.* Highly Efficient Slow-Light Mach–Zehnder Modulator Achieving 0.21 V cm

Efficiency with Bandwidth Surpassing 110 GHz. *Laser & Photonics Reviews* **19**, 2401092, (2025).

37. A. Melikyan, K. Kim, B. Stern, and N. Kaneda, "Self-biasing of carrier depletion based silicon microring modulators," *Opt. Express,* vol. 28, no. 15, pp. 22540-22548, 2020/07/20 2020.

38. Müller, J. et al. Optical peaking enhancement in high-speed ring modulators. Sci. Rep. 4, 6310 (2014).

39. Karimelahi, S. & Sheikholeslami, A. Ring modulator small-signal response analysis based on pole-zero representation. Opt. Express 24, 7585–7599 (2016).

40. P. Dong, S. Liao, D. Feng, H. Liang, D. Zheng, R. Shafiiha, C. C. Kung, W. Qian, G. Li, X. Zheng, A. V. Krishnamoorthy, and M. Asghari, "Low V pp, ultralow-energy, compact, high-speed silicon electro-optic modulator," Opt. Express **17**(25), 22484–22490 (2009).

41. G. L. *et al.*, "Ring resonator modulators in silicon for interchip photonic links," *IEEE J. Sel. Topics Quantum Electron.*, vol. 19, no. 6, pp. 95–113, Nov./Dec. 2013

42. David W. U. Chan and Hon Ki Tsang, "Sub-volt forward-biased silicon microring modulator at 210 Gb/s," Opt. Lett. 49, 6477-6480 (2024)

43. C. S. Levy et al., "8-λ × 50 Gbps/λ Heterogeneously Integrated Si-Ph DWDM Transmitter," IEEE Journal of Solid-State Circuits, vol. 59, no. 3, pp. 690-701, 2024.

44. D. D. Sharma, G. Pasdast, Z. Qian, and K. Aygun, "Universal Chiplet Interconnect Express (UCIe): An Open Industry Standard for Innovations With Chiplets at Package Level," IEEE Transactions on Components, Packaging and Manufacturing Technology, vol. 12, no. 9, pp. 1423-1431, 2022.

45. "GlobalFoundries 12 LP—CMC microsystems," https://www.cmc.ca/globalfoundries-12-lp/.

46. Y. Zhang et al., "36.6 A 112Gb/s 0.61pJ/b PAM-4 Linear TIA Supporting Extended PD-TIA Reach in 28nm CMOS," in 2025 IEEE International Solid-State Circuits Conference (ISSCC), 2025, vol. 68, pp. 1-3.

47. PCI Express® Base Specification Revision 5.0 Version 1.0, https://picture.iczhiku.com/resource/eetop/SYkDTqhOLhpUTnMx.pdf

48. Pegah Alavi, et al., "PCI Express & PAM4: The Pathway to 128GT/s and Challenges of Building Interoperable 64GT/s Capable Systems", in DesignCon, 2024.

**Acknowledgements:**

This work is supported by Zhangjiang Laboratory.

**Author contributions:**

The concept of this work was conceived by F.H., F.Y., X.H. and W.C. The device design was performed by F.Y. and X.H. The device characterization and high-speed experiments were performed by F. H., F.Y., X.L. and A.W. The results were analyzed by F.H., F.Y., X.H and W.C. All authors participated in the writing of the manuscript. The project was under supervision of X.H., H.C. and W.C.

**Competing interests:**

The authors declare no competing interests.

**Data availability:**

All the data generated in this study are available from the corresponding author upon reasonable request.

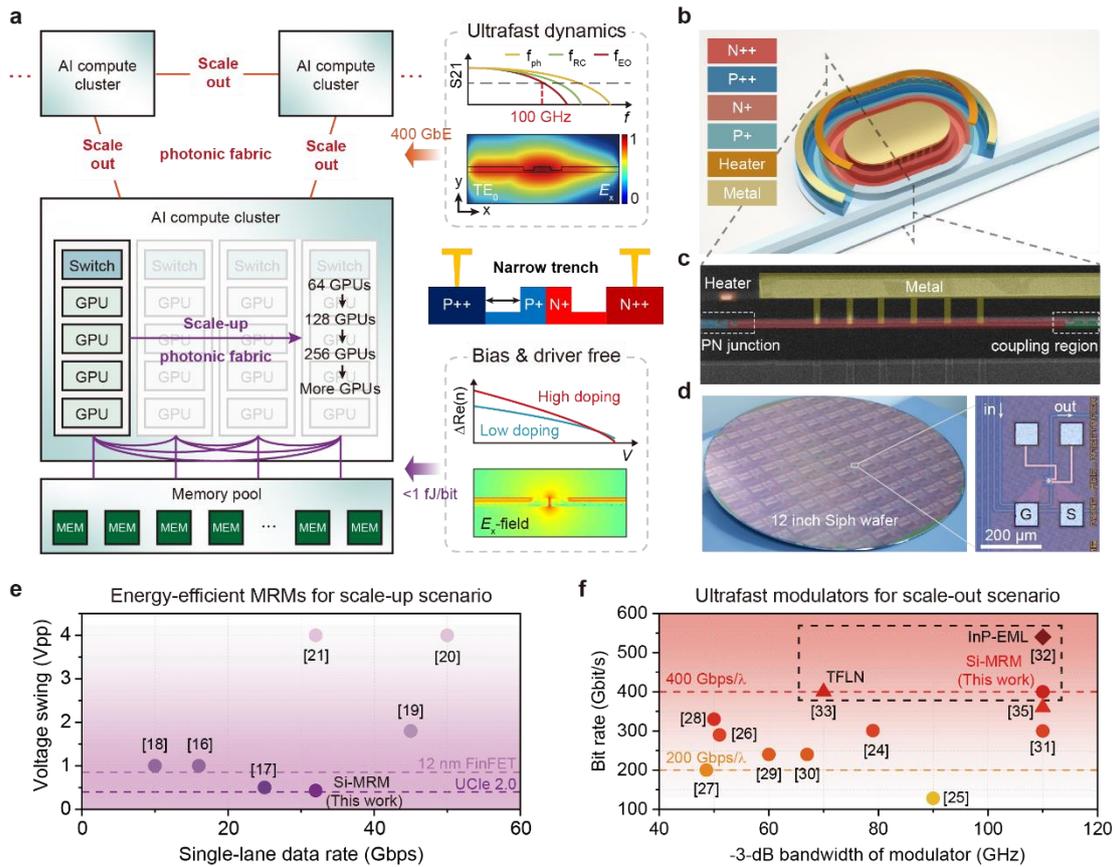

**Fig.1 Schematic diagram of our proposed ultrafast silicon microring modulators (MRMs) enabling the XPUs scale-up and scale-out in AI computing infrastructure. a.** Conceptual illustration of AI compute networks, where the photonic fabric (optical interconnects) links GPUs within or across racks and connects different compute clusters. Our microring modulator (MRM), featuring ultrafast dynamics and a bias- & driver-free operation, provides a unified route toward scalable interconnects: enhancing the scale-up of processing units (XPU) and the scale-out of compute clusters. **b.** 3D schematic of the MRM. **c.** Cross-sectional FIB-SEM image of the full device. **d.** Photograph of a 12-inch silicon photonics wafer and a microscopic close-up of the MRM. **e.** Performance benchmarking shows that our device achieves high single-lane data rates at low RF voltage swings, highlighting its suitability for energy-efficient, error-free XPU scale-up. **f.** The device further delivers record-high bit rates (400-Gbps-per-λ) with large modulation bandwidths, underscoring its potential for high-capacity Ethernet scale-out interconnects.

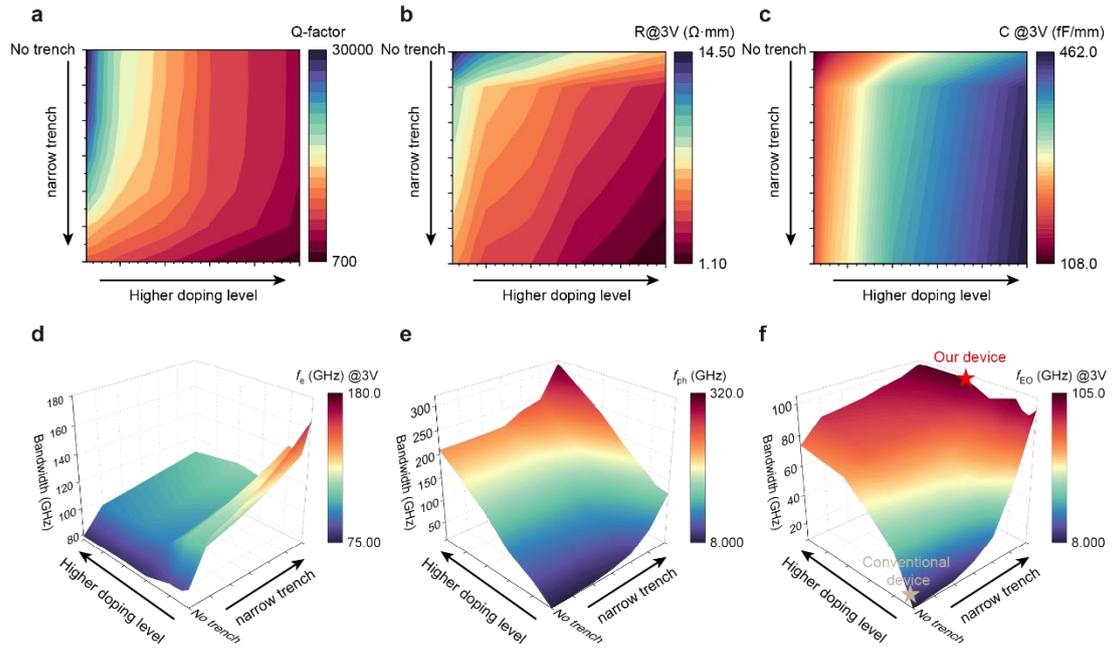

**Fig.2 Design and simulation of the MRM.** The dependence of **a.** Q-factor, **b.** resistance (*R*) and **c.** capacitance (*C*) on trench width and doping concentration in the MRM. The dependence of **d.** electrical bandwidth ($f_e$), **e.** optical bandwidth ($f_{ph}$) and **f.** overall electro-optical bandwidth ($f_{EO}$) on trench width and doping concentration. All simulation of the MRM is operated at the -3V electrical bias. The simulation result of the optimized MRM parameters is marked in (f) via a red star-shaped marker, showing $f_{EO}$ over 100 GHz.

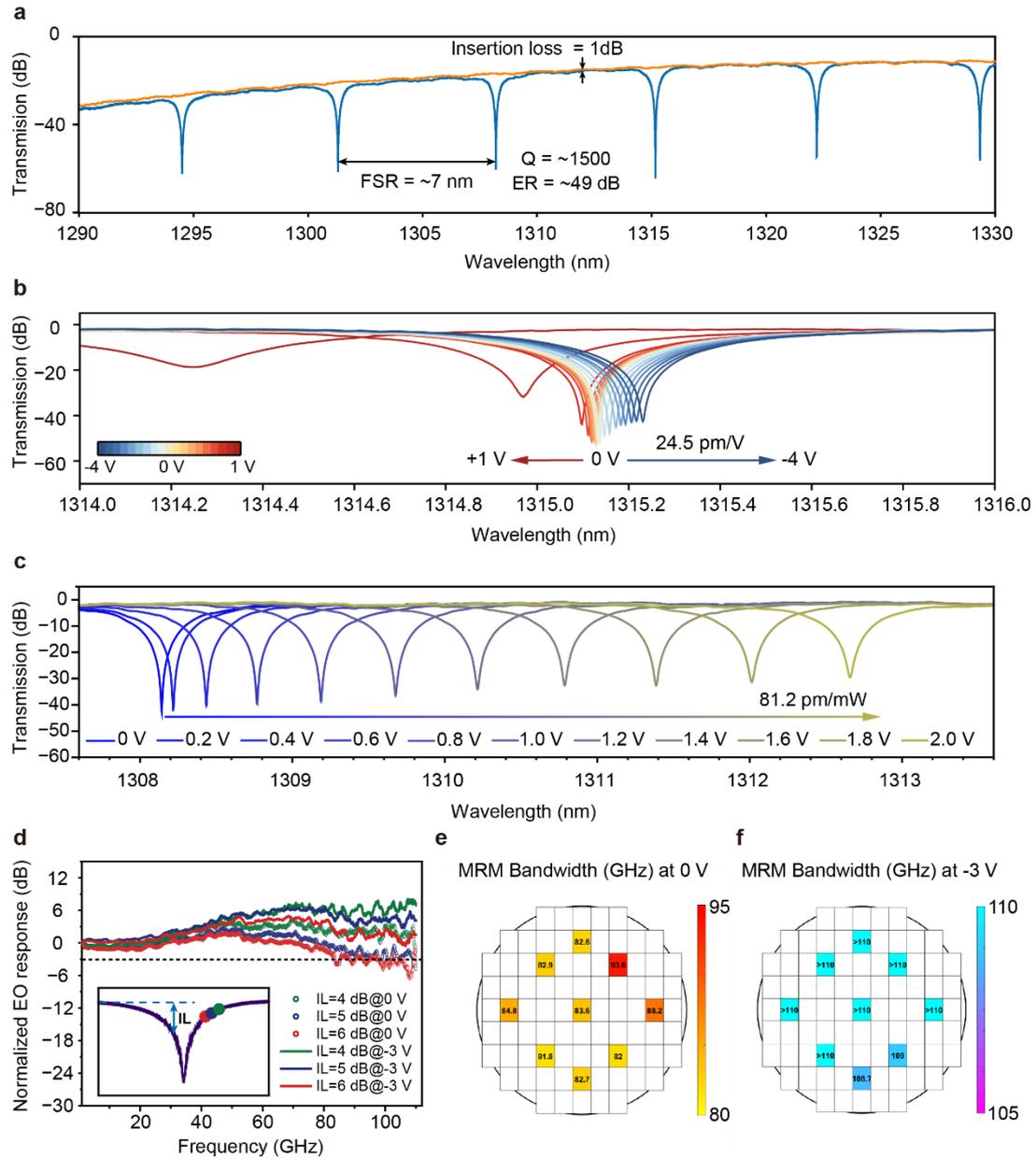

**Fig.3 Direct current (DC) characterization and wafer-scale frequency response of the MRM.**

**a.** Measured transmission spectrum of the MRM (blue) and a pair of grating couplers (orange). **b.** Measured transmission spectrum versus junction voltage from 1 V to -4 V. **c.** Measured transmission spectrum versus heater voltage ranging from 0 V to 2 V. **d.** Measured electro-optic (EO) responses with insertion loss from 4 to 6 dB, at bias voltage of 0 and −3V, respectively. **e,f.** Distribution of the -3 dB bandwidth on 12-inch wafer at 0 and -3 V and insertion loss of 6 dB.

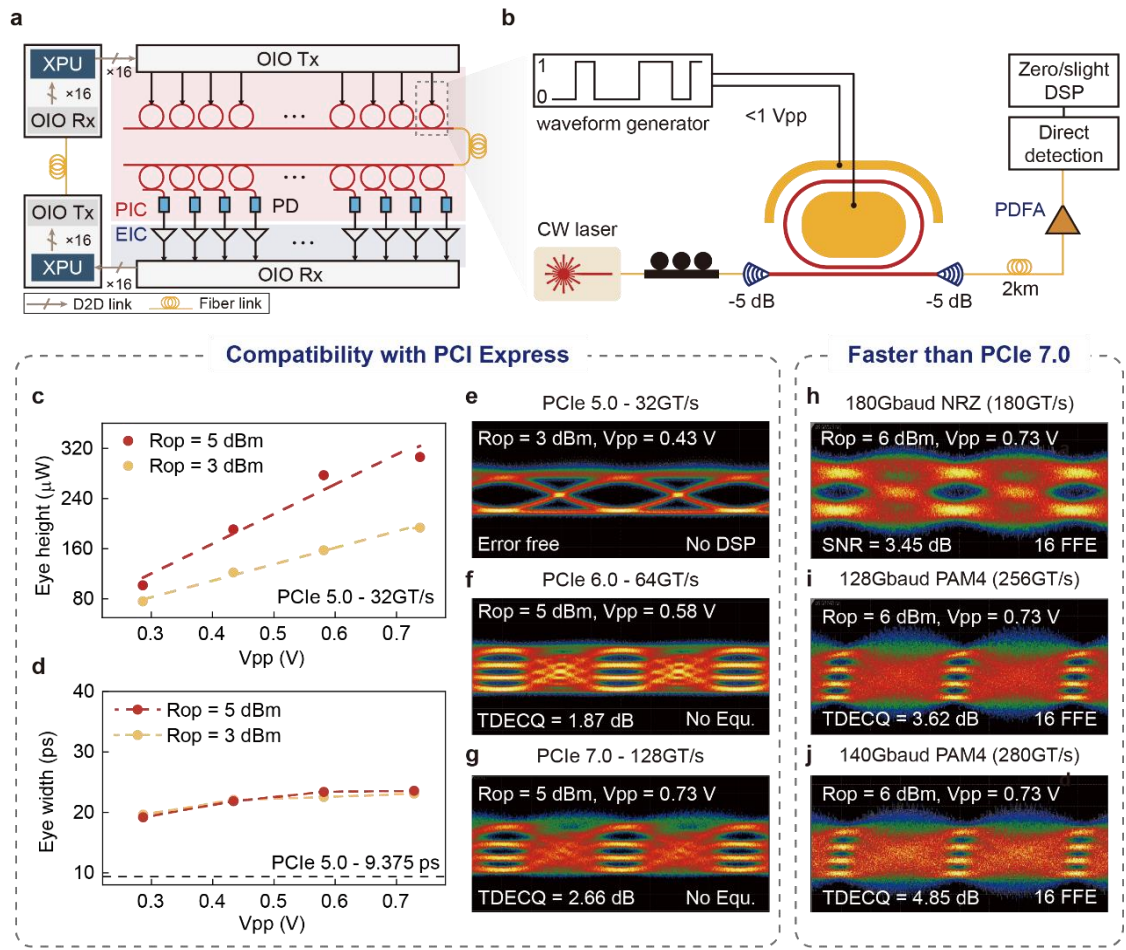

**Fig. 4. Demonstration of an electrical-bias-free and driver-free MRM as a PCIe-compatible optical transmitter for XPU interconnection scale-up. a.** Schematic illustration of XPU-to-XPU optical I/O (OIO) interconnects, where die-to-die (D2D) electrical links interface with the OIO and comply with PCIe standards. The OIO Tx and Rx consist of Serializer/Deserializer (SerDes) and phase lock loop (PLL) circuit. **b.** Experimental setup for eye diagram measurement using a single MRM driven with sub-1 Vpp signals, without electrical bias or driver. Minimal or no digital signal processing (DSP) is applied at the receiver. **c-g.** PCIe physical-layer compliance test results obtained with the MRM-based transmitter. **h-j.** Measured eye diagrams of the MRM operating at data rates exceeding those of PCIe 7.0.

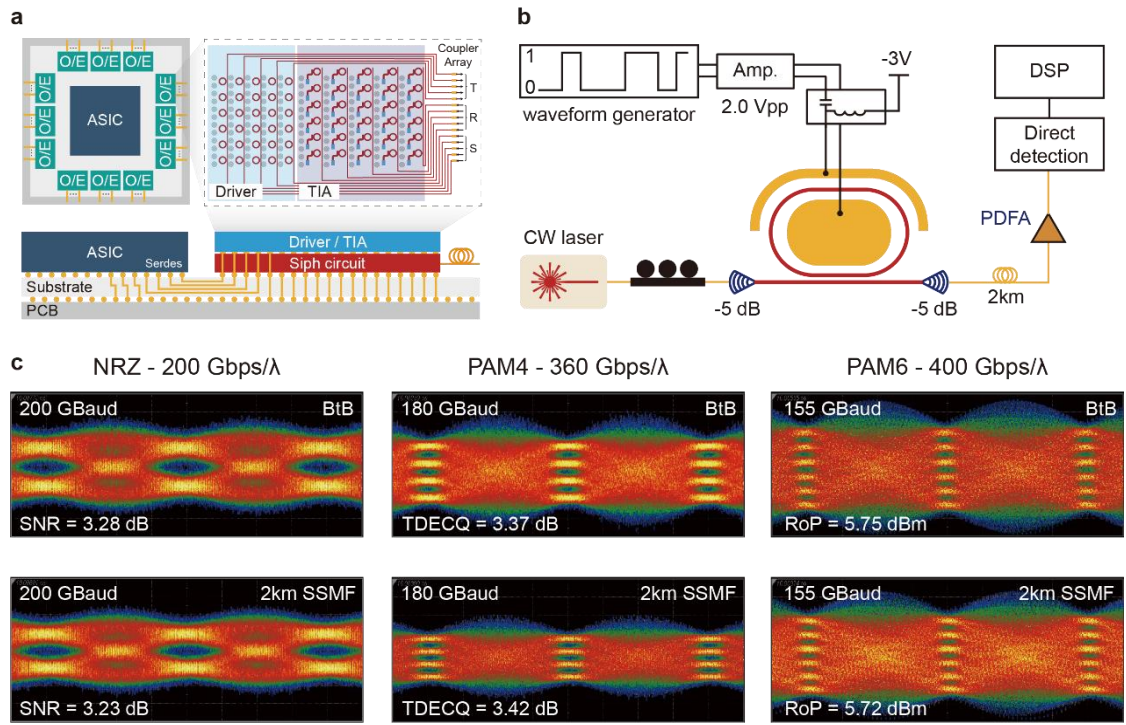

**Fig. 5. Demonstration of an ultrafast MRM as a 400Gbps/λ optical transmitter for switch interconnects scale-out. a.** Schematic illustration of high-speed optical interconnects enabled by MRM-based optical transceiver for Ethernet switch. **b**. Experimental setup of a bias- and driver-assisted MRM for ultrahigh-speed transmission, where the driving signal is amplified to 1 Vpp and applied to the device under test (DUT). Digital signal processing (DSP) is used at the receiver to mitigate inter-symbol interference. **c.** Measured eye diagrams of the MRM operating at data rates up to 400 Gbps/λ for back-to-back (BtB) and 2km standard single mode fiber (SSMF).

# Supplementary Information for

# 400-Gbps/λ Ultrafast Silicon Microring Modulator for Scalable Optical Compute Interconnects

Fangchen Hu[1,3], Fengxin Yu[1,3], Xingyu Liu[2], Aoxue Wang[2], Xiao Hu[1,*], Haiwen Cai[1,*], and Wei Chu[1,*]

[1]Zhangjiang Laboratory, Shanghai 201210, China

[2] Future Information Innovative College, Fudan University, Shanghai 200433, China

[3]These authors contribute equally.

[*]Corresponding author. E-mail: huxiao@zjlab.ac.cn (X.H.), caihw@zjlab.ac.cn (H.C.) and chuwei@zjlab.ac.cn (W.C.)

# Content

**Supplementary Note 1: Simulation and design of MRM**

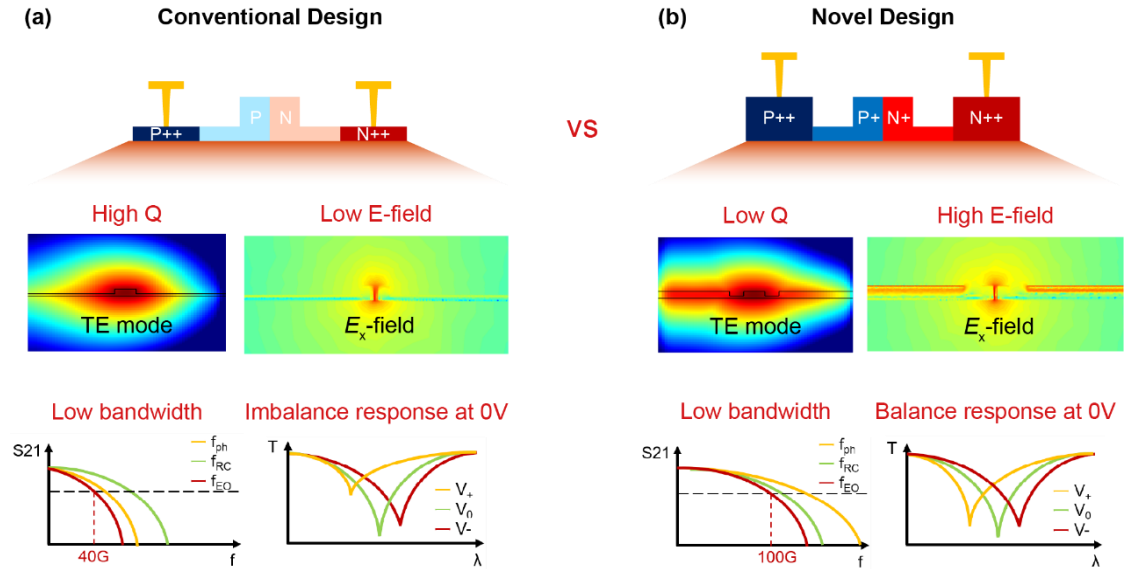

**Fig. S1. Comparison of the novel design and conventional design. a.** Conventional design of the MRM. **b.** Our proposed design.

**Fig. S1** compares a conventional PN-junction-based microring modulator (MRM) with our newly designed symmetric-junction MRM in terms of optical field distribution and modulation response. In the conventional asymmetric PN-junction structure, the resonator exhibits a high optical Q-factor but suffers from weak overlap between the optical mode and the applied electric field. As a result, the modulation efficiency is low, the electro-optic bandwidth is restricted to below 40 GHz, and the transmission spectrum at 0 V bias becomes imbalanced, which degrades eye quality under bias-free operation.

By contrast, the proposed symmetric P+N+ junction configuration reduces the Q-factor while significantly enhancing the optical–electrical field overlap. This leads to higher modulation efficiency and extends the 3-dB electro-optic bandwidth to beyond 100 GHz, effectively mitigating the RC-limited response observed in conventional

MRMs. Moreover, the transmission spectra remain balanced at 0 V bias, ensuring stable and distortion-free zero-bias operation.

Overall, this comparison underscores the fundamental trade-off between Q-factor and modulation efficiency, and demonstrates that a symmetric junction design can simultaneously deliver large bandwidth and robust bias-free operation—both of which are essential for realizing energy-efficient high-speed optical interconnects.

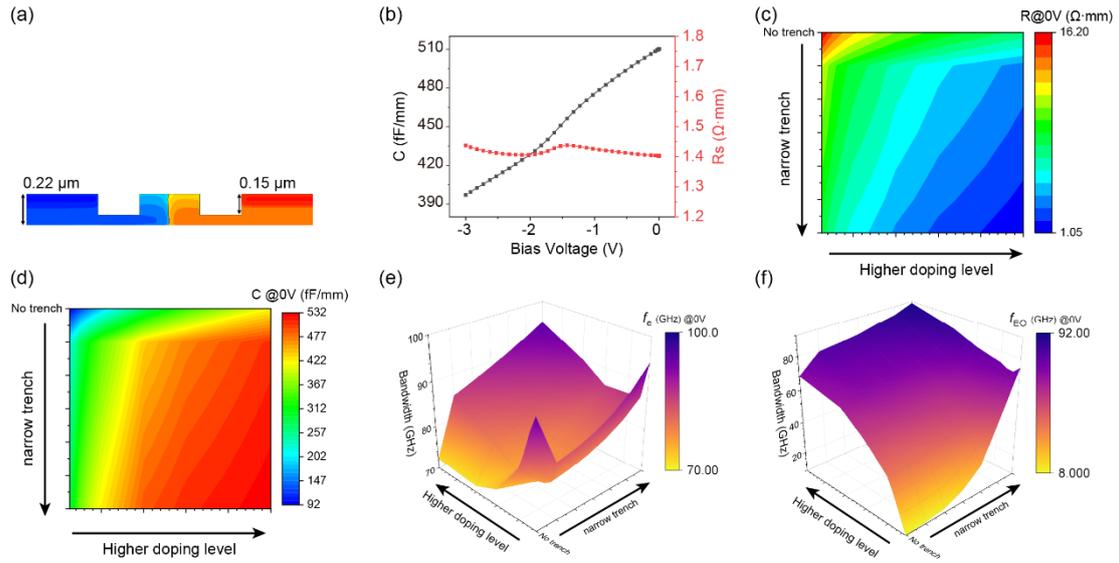

**Fig. S2. Simulated doping concentration, resistivity and capacitance of MRM.** (a) Simulated doping distribution of the MRM. (b) Simulated junction capaticance and series resistance per unit length of the target MRM versus bias voltage. The dependence of (c) series resistance, (d) junction capaticance, (e) electrical bandwidth and (f) electro-optical bandwidth on trench width and doping concentration.

Based on the ion implantation condition and annealing process during real fabrication, Synopsys Sentaurus Process was used to obtain the simulated doping distribution of the microring modulator (MRM). The simulated doping profile of the optimized device is shown in **Fig. S2**(a). The average p-type and n-type doping are located in the 420nm-wide, 220 nm-thick Si waveguide with a 70nm-thick slab. A device simulation was then performed with the Synopsys Sentaurus Device tool to obtain the distribution of free carriers in the waveguide for different bias voltages. Importing the doping profile into Sentaurus Device enabled extraction of p-n junction capacitance and series resistivity values via small-signal AC analysis. The simulated junction capaticance and series resistance per unit length of the target MRM versus bias voltage are plotted in Fig. S2(b). Combined with the length (40 μm) of the doped region in microring, the p-n junction has a capacitance $C_j$ of ~21 fF/16 fF and a series resistance $R_s$ of ~31 Ω/ 32 Ω at 0 V and -3 V respectively. To examine how doping and trench width affect MRM junction capacitance and series resistivity, we plotted both parameters versus these variables at 0 V (Figure S2(c-d)) and -3 V. It is obvious that the capacitance decreases as the trench width increases and doping concentration decreases. The series resistivity decreases as the trench width decreases and doping concentration increases.

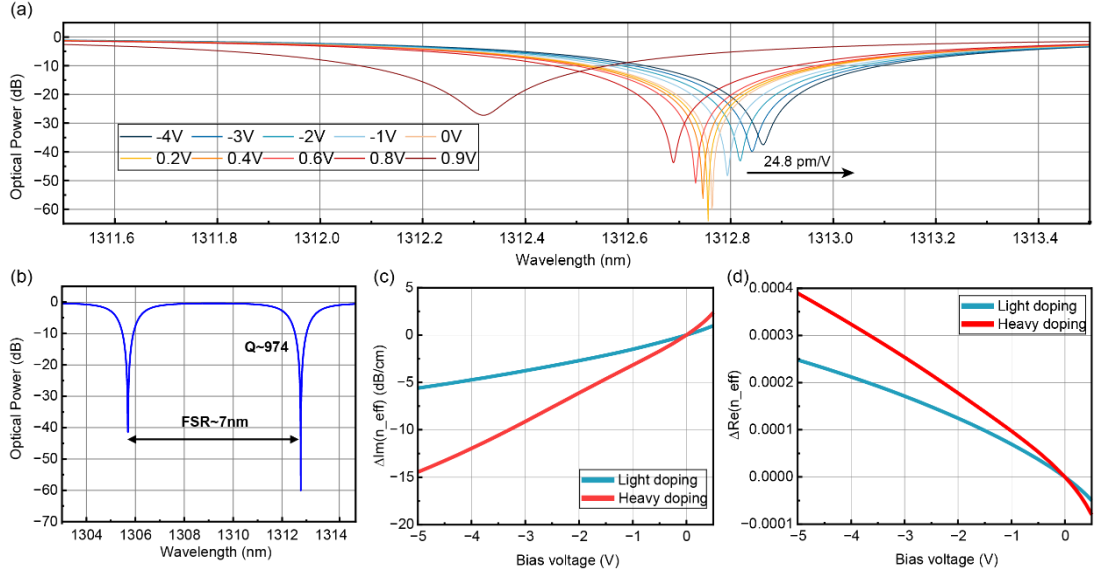

**Fig. S3 Transmission spectrum, mode effective refractive Index of MRM under bias voltage.**

(a) Simulated transmission spectra under -4 to +0.9 V bias. (b) Spectrum at 0 V bias. (c-d) Simulated $\Delta Re(n_{\text{eff}})$ (c) and $\Delta Im(n_{\text{eff}})$ (d) under bias: our design (red) vs. conventional rings (blue)

In an all-pass microresonator with a ring coupled to a bus waveguide, under single-mode and steady-state conditions, the transmission spectrum is given by:

$$T = \left|\frac{E_{out}}{E_{in}}\right|^2 = \frac{a^2+\tau^2-2a\tau\cos(\varphi)}{1+a^2\tau^2-2a\tau\cos(\varphi)} \quad (1)$$

where $\tau$, $\varphi$ and $a$ are field transmission coefficient of the coupler, round-trip phase shift, and amplitude transmission, respectively. For lossless coupling, the cross-coupling coefficient ($\kappa$) and through-coupling coefficient ($\tau$) satisfy $|\tau|^2 + |\kappa|^2 = 1$. The phase shift $\varphi$ depends on the bent waveguide's effective index and microring circumference. The microring's intrinsic loss (1-$a^2$), arising from bending, material absorption, and scattering, is directly linked to the imaginary component of the waveguide mode's effective refractive index.

Lumerical simulations of the microring transmission spectra enabled extraction of key parameters: quality factor (Q), free spectral range (FSR), and phase shift efficiency.

Lumerical FDTD simulations quantified gap-dependent coupling coefficients between the bus and microring, while MODE's eigenmode analysis and CHARGE's electrostatics solver jointly characterized the bent waveguide's voltage-tunable effective refractive index. These results were then imported into INTERCONNECT to model the microring modulator and simulate its transmission spectra under various bias voltages.

As shown in **Fig. S3**(b), the microring has simulated 7 nm FSR and Q-factor of 974, lower than measured Q-factor (1500). The difference may come from the overestimation of the waveguide loss and different coupling conditions. **Fig. S3**(a) shows the simulated transmission spectra with different bias voltages applied, from −4 V to +0.9 V. The average wavelength shift reaches 24.8 pm/V as the bias voltage varies from 0 to -4 V, matching well with result (24.5 pm/V) obtained from 12-inch wafer-level measurement.

**Figs. S3**(c)-(d) compare the effective index changes ($\Delta n_{eff}$) between our novel microring and conventional low-doped structures under different bias. Our design shows larger $\Delta n_{eff}$ and phase-shift efficiency than low-doped rings at the same bias. Despite exhibiting larger loss fluctuations, the device's inherently high intrinsic loss suppresses variations in coupling conditions.

**Supplementary Note 2: Device characterization on a 12-inch wafer**

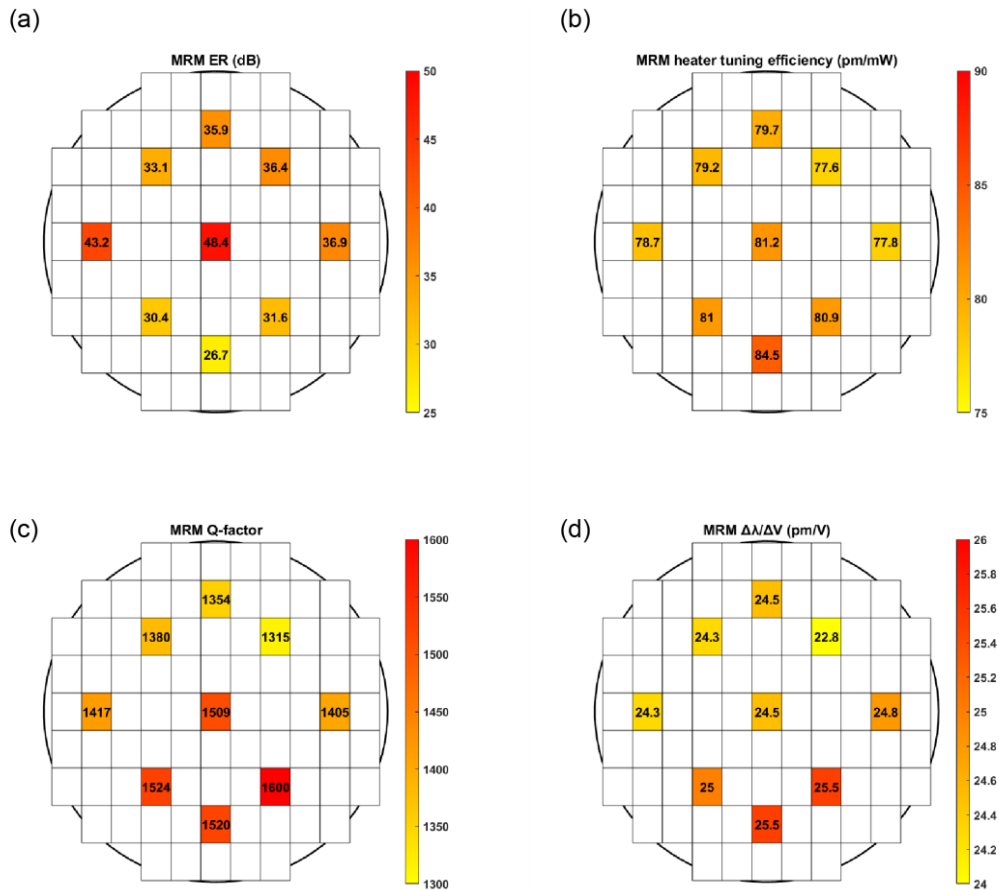

**Fig. S4 Measured results of ER, heater tuning efficiency, Q-factor and electrical modulation efficiency on a 12 inch wafer.** (a) ER. (b) Heater tuning efficiency. (c) Q-factor. (d) electrical modulation efficiency ($\Delta\lambda/\Delta V$).

**Fig. S4** summarizes the measured characteristics of our fabricated MRMs across a full 12-inch wafer, including extinction ratio (ER), heater tuning efficiency, Q-factor, and electrical modulation efficiency. The ER of the devices ranges from 26.7 dB to 48.4 dB, indicating strong on–off contrast and uniform modulation quality across the wafer [**Fig. S4**(a)]. The measured heater tuning efficiency is between 77.6 and 84.5 pm/mW [**Fig. S4**(b)], demonstrating stable thermal tuning behavior. The intrinsic optical Q-factor lies in the range of 1315 to 1600 [Fig. S4(c)], confirming low excess loss introduced during fabrication. Finally, the electrical modulation efficiency ($\Delta\lambda/\Delta V$) is measured between 22.8 and 25.5 pm/V [Fig. S4(d)], highlighting the reproducibility of junction doping and electrode design. These results validate the uniformity and reliability of our wafer-scale fabrication process, ensuring scalability for large-volume silicon photonics integration.

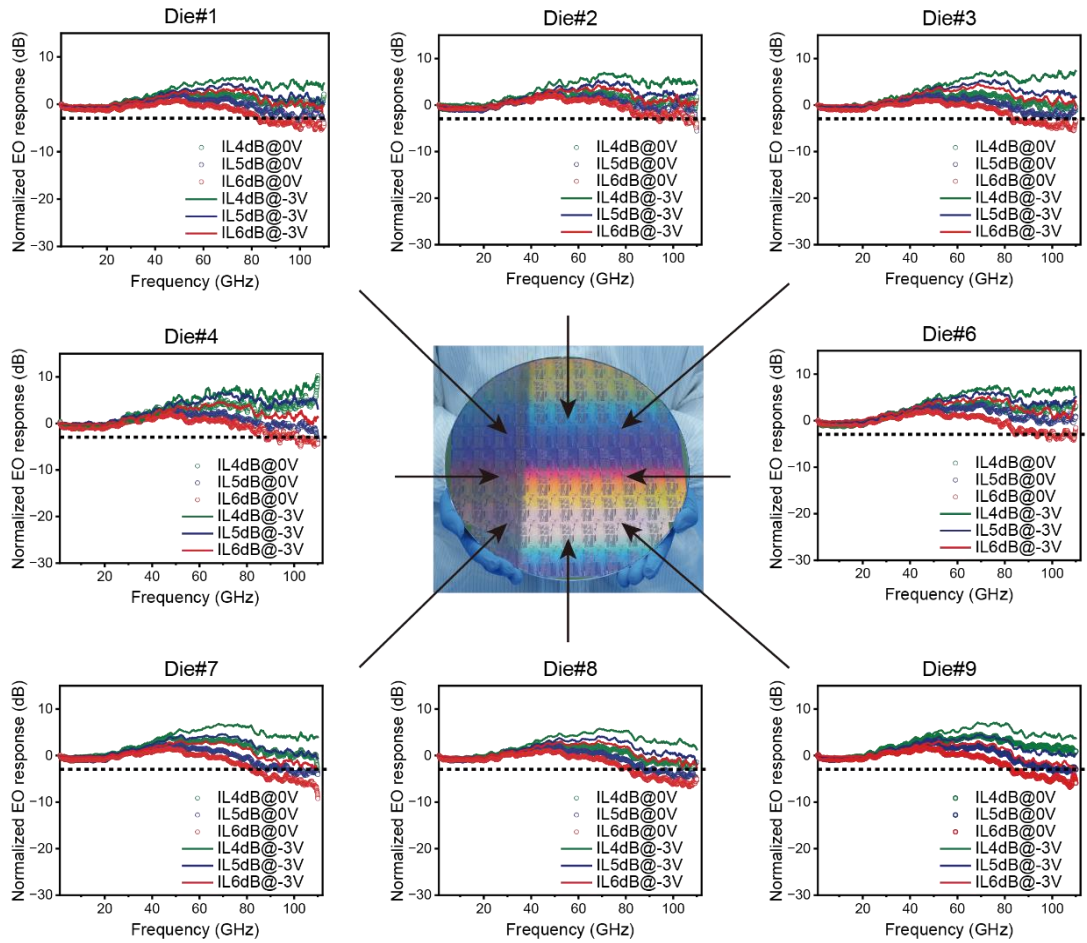

**Fig. S5 Bandwidth Uniformity.** Measured bandwidth of MRMs at different IL and electrical bias on different dies of a 12 inch wafer.

**Fig. S5** provides detailed characterization of the bandwidth uniformity of the microring modulators (MRMs) across multiple dies on a 12-inch wafer. The figure presents measured S21 curves at various operating points, including different insertion losses (IL) and electrical bias conditions, for nine representative dies. The data demonstrate the exceptional consistency of the modulators' bandwidth, highlighting the feasibility of commercial-scale production and reliable high-speed operation across the wafer. The die #5 is shown in Fig. 3d of the main text.

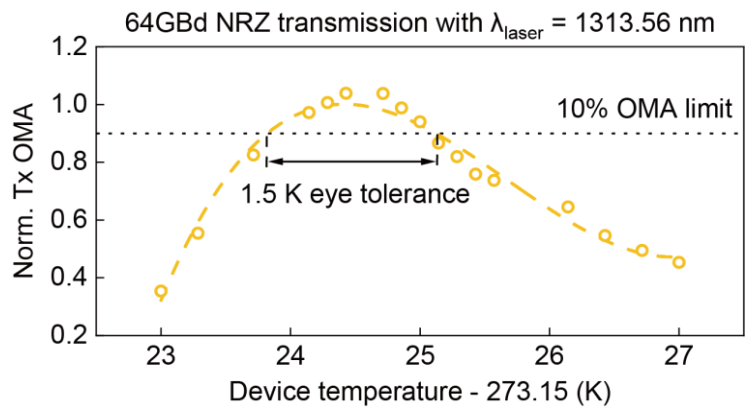

**Fig. S6. Temperature sensitivity of our MRM.** Plot showing the normalized OMA (eye opening) of a 64 Gbaud NRZ signal transmitted with electrical bias-free on a fixed-wavelength carrier (1313.56 nm) plotted over the device temperature. The 1.5 K operation range is clearly visible.

**Supplementary Note 3: High-speed transmission performance**

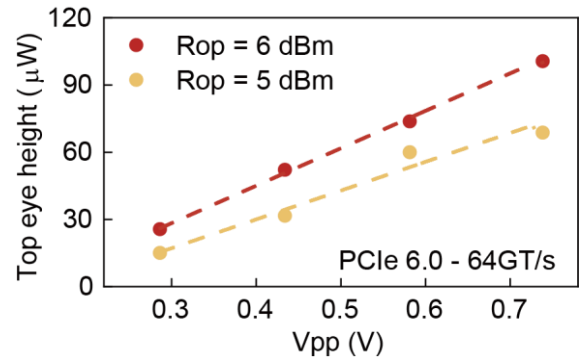

**Fig. S7. Optical eye height of PCIe 6.0 signal modulated on the MRM.** The top eye height of PAM 4 signal refers to the vertical distance between the highest voltage level and the adjacent lower level in the eye diagram.

**Fig. S7** presents the optical eye diagram analysis for PCIe 6.0 signals, extending the evaluation of our microring modulator (MRM) architecture to higher data rates. Similar to the PCIe 5.0 measurements, the experiments were performed by sweeping the RF drive voltage (Vpp) below the UCIe 2.0 specification and the 12 nm FinFET core voltage, ensuring tight voltage compatibility with modern CMOS systems. The figure shows the resulting optical eye heights and eye widths as functions of Vpp and optical received power (Rop). The measured eye metrics demonstrate that the modulator can support PCIe 6.0 lane rates while maintaining sufficient optical eye opening, providing practical guidance for selecting transimpedance amplifiers (TIAs) and variable gain amplifiers (VGAs) to satisfy electrical eye height requirements.

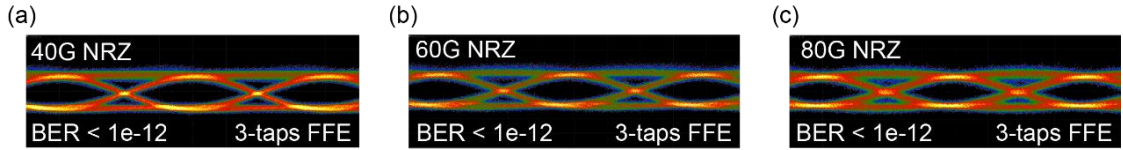

**Fig. S8. Optical eye diagrams of NRZ signal modulated on the self-biasing MRM.** When applying 3-taps FFE, BERs of (a). 40 Gbaud, (b). 60 Gbaud and (c). 80 Gbaud signal are all lower than 1e-12, which means up to 80 Gbps signal can be transmitted without the use of FEC. The Rop and $V_{pp}$ is 6 dBm and 0.56 V, respectively.

**Fig. S8** shows the optical eye diagrams of NRZ signals modulated on the self-biasing microring modulator (MRM) at different baud rates. When a 3-tap feed-forward equalizer (FFE) is applied, the measured bit error rates (BERs) for 40 Gbaud, 60 Gbaud, and 80 Gbaud signals are all below $1\times10^{-12}$, indicating that data rates up to 80 Gbps can be transmitted without forward error correction (FEC). The measurements were conducted with an optical received power (Rop) of 6 dBm and an RF drive voltage (Vpp) of 0.56 V, demonstrating the high-speed transmission capability of the self-biasing MRM.

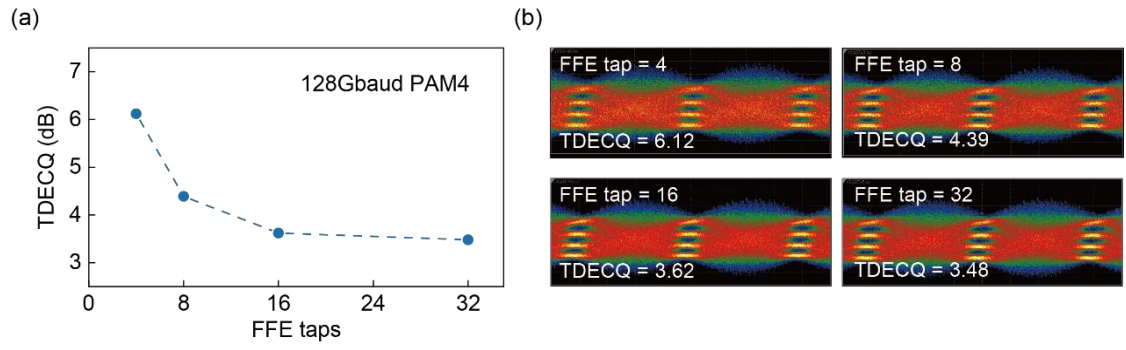

**Fig. S9. Impact of FFE taps on the TDECQ of a 128 Gbaud PAM4 signal.** (a) TDECQ as a function of the number of FFE taps. (b) Corresponding eye diagrams at different FFE taps.

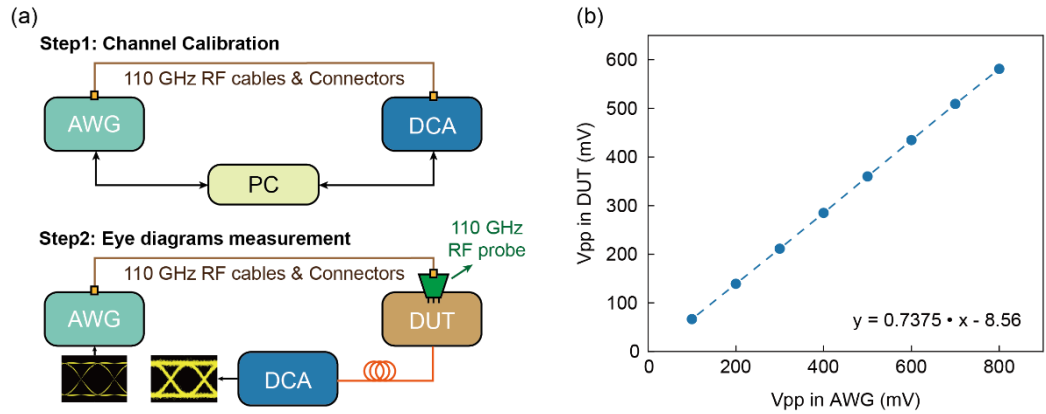

**Fig. S10. Calibration of the test systems for eye diagram measurement and RF peak-to-peak voltage characterization on the devices under test (DUTs).** (a) In the first step, the frequency response of the entire channel—including the AWG, DCA, and all RF cables and connectors—is flattened over the 0 - 90 GHz range. After calibration, the DUT is connected to the AWG and DCA via 110 GHz ground–signal (GS) RF probes and optical fibers, respectively. This calibrated setup enables accurate eye diagram measurement on the DCA. The AWG provides RF signals with tunable peak-to-peak voltage ($V_{pp}$), modulation formats, and baud rates. (b) Relationship between the actually measured $V_{pp}$ delivered to the DUT and the nominal Vpp set on the AWG. The dotted line represents a linear fit to the measured data.

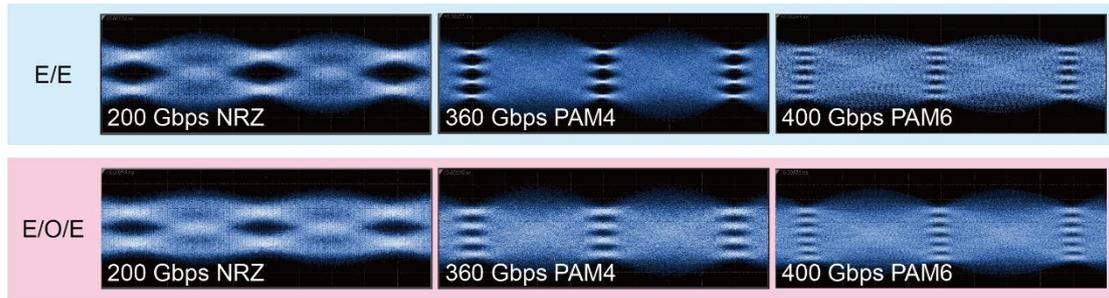

**Fig. S11. Comparison of eye diagrams measured at E/E and E/O/E configurations.** In the E/E case, the RF electrical signal is directly fed into the electrical input of the sampling oscilloscope. In the E/O/E case, the electrical signal is first modulated by the microring modulator (MRM) as an optical transmitter, and then the optical output is received by the optical input of the oscilloscope.

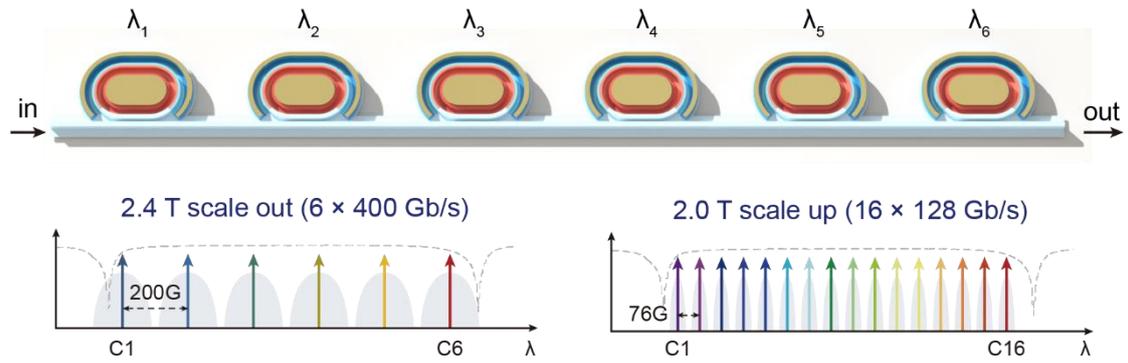

**Fig. S12. Scalability of this MRM for WDM interconnection.** By cascading 6 depletion-mode MRMs in a straight bus waveguide, the WDM architecture can support 2.4 Tbps interconnection for scale out scenarios. In this architecture, the channel spacing of each ring is set to 200 GHz to accommodate 155Gbuad PAM6 (400 Gbps) within a FSR range of ~7 nm. By cascading 16 self-biasing-mode MRMs in a straight bus waveguide, the WDM architecture can support 2 Tbps interconnection for scale up scenarios. The channel spacing of each ring is set to 76 GHz to accommodate 64 Gbaud PAM4 (128 Gbps).

**Supplementary Table 1:** Literature overview of high-efficiency microring modulators for error-free scale-up scenarios

| Year | Ref | Speed (Gbps) | Efficiency (fJ/bit) | Vpp (V) | Capacitance (fF) | BER | Operation mode |
|---|---|---|---|---|---|---|---|
| 2014 | [1] | 25 | 0.9 | 0.5 | 17 | <1E-12 | Depletion-mode |
| 2023 | [2] | 16 | 12.5 | 1 | 50 | <1E-12 | Depletion-mode |
| 2024 | [3] | 45 | \ | 1.8 | \ | <1E-12 | Depletion-mode |
| 2024 | [4] | 32 | 1.87 | 4 | \ | <1E-12 | Depletion-mode |
| 2025 | [5] | 50 | \ | 4 | \ | <1E-12 | Depletion-mode |
| 2025 | This work | 32 | 0.97 | 0.43 | 21 | <1E-12 | Self-biasing mode |

**Supplementary Table 2:** Literature overview of high-speed mircroring modulators of P-N junction.

| Type | Ref | Radius (μm) | ER (dB) | FSR (nm) | Q factor | $V_\pi L$ (V·cm) | Electrical tuning (pm/V) | Bandwidth (GHz) | operating wavelength | Line Rate (Gbps) | Applied $V_{pp}$ (V) |
|---|---|---|---|---|---|---|---|---|---|---|---|
| Si-RT | This work | 8 | 49 | 7 | 1500 | 0.57 | 24.5 | 110 | O-band | 400 (PAM6) 360 (PAM4) 200 (NRZ) | 2 |
| Si-RT | [6] | 8 | 30 | 7 | 8000 | 0.825(-4.5V) | 26.4 | 79 | O-band | 301 (DMT) 240 (PAM8) 200 (PAM4) 120 (NRZ) | 3 |
| Si-MRM | [7] | 4 | 18 | 22.9 | 2670 | / | 30.4 | 90 | C-band | 128 (NRZ) 200 (PAM4) | 1.36 |
| Si-MRM | [83] | 5 | 23 | 13 | 26300 | 0.9 | 15 | 51 | O-band | 290 (PAM6) 224 (PAM4) | 0.4 |
| Si-MRM | [9] | 12 | 16 | 5.7 | 3700 | 0.6 | 27.3 | 48.6 | O-band | 200 (PAM4) | 1.6 |
| Si-MRM | [10] | 7.5 | 24 | / | 5200 | 0.63 | / | 50 | C-band | 330 (PAM8) | 1.8 |
| Si-MRM | [11] | 4 | 19 | 16.3 | 4000 | 0.42 | / | 60 | O-band | 240 (PAM4) | 1.8 |
| Si-MRM | [12] | 3 | 16.5 | 35.7 | 1170 | 1.35 | 25 | >110 | C-band | 300 (PAM4) | 2 |

*Si-MRM: silicon microring modulator; Si-RT: Silicon micro-racetrack modulator.

**References:**

[1] Timurdogan, E. et al. An ultralow power athermal silicon modulator. *Nature Communications* 5, 4008, (2014).

[2] Rizzo, A. *et al.* Massively scalable Kerr comb-driven silicon photonic link. *Nature Photonics*, doi:10.1038/s41566-023-01244-7 (2023).

[3] C. Xie *et al.*, "A 64 Gb/s NRZ O-Band Ring Modulator with 3.2 THz FSR for DWDM Applications," in *2024 Optical Fiber Communications Conference and Exhibition (OFC)*, 2024, pp. 1-3.

[4] C. Sun, "Photonics for Die-to-Die Interconnects: Links and Optical I/O Chiplets", in ISSCC 2024, Forum 1.7.

[5] N. Qi et al., "A Monolithically Integrated DWDM Si-Photonics Transceiver for Chiplet Optical I/O," IEEE Journal of Solid-State Circuits, pp. 1-13, 2025.

[6] Y. Zhang et al., "240 Gb/s optical transmission based on an ultrafast silicon microring modulator," Photon. Res., vol. 10, no. 4, pp. 1127-1133, 2022/04/01 2022.

[7] S. Zhao et al., "High Bandwidth and Low Driving Voltage Add-Drop Micro-Ring Modulator for Optical Interconnection I/O Chips," Journal of Lightwave Technology, pp. 1-9, 2025.

[8] X. Wang et al., "A 290 Gbps Silicon Photonic Microring Modulator with 83 -aJ/bit Power Consumption," in Optical Fiber Communication Conference (OFC) 2025, San Francisco, California, 2025: Optica Publishing Group, p. M3K.5.

[9] Y. Yuan et al., "A 5 × 200 Gbps microring modulator silicon chip empowered by two-segment Z-shape junctions," Nature Communications, vol. 15, no. 1, p. 918, 2024/01/31 2024.

[10] D. W. U. Chan, X. Wu, C. Lu, A. P. T. Lau, and H. K. Tsang, "Efficient 330-Gb/s PAM-8

modulation using silicon microring modulators," Opt. Lett., vol. 48, no. 4, pp. 1036-1039, 2023/02/15 2023.

[11] M. Sakib et al., "A 240 Gb/s PAM4 Silicon Micro-Ring Optical Modulator," in 2022 Optical Fiber Communications Conference and Exhibition (OFC), 2022, pp. 01-03.

[12] K. Lu, H. Chen, W. Zhou, H. K. Tsang, and Y. Tong, "Whispering Gallery Mode Enhanced Broadband and High-Speed Silicon Microring Modulator," Journal of Lightwave Technology, pp. 1-7, 2025.